\documentclass[12pt,psfig]{article}
\usepackage{amssymb}
\usepackage{amsmath}
\usepackage{a4wide}
\usepackage{epsfig}
\def\t{\tilde}
\def\be{\begin{equation}}
\def\ee{\end{equation}}
\def\bea{\begin{eqnarray}}
\def\eea{\end{eqnarray}}

\def\f#1#2{\frac{#1}{#2}}

\begin{document}
\begin{titlepage}
\title{{\bf LTB solutions in Newtonian gauge: from strong to weak
fields. }} \vskip2in

\author{
{\bf Karel Van Acoleyen$$\footnote{\baselineskip=16pt E-mail: {\tt
karel.vanacoleyen@ugent.be}}}
\hspace{3cm}\\
 $$ {\small Department of Mathematical Physics and Astronomy, Ghent
 University,}\\
 {\small Krijgslaan 281, S9, 9000 Gent, Belgium}.
}

\date{}
\maketitle
\def\baselinestretch{1.15}
\begin{abstract}
\noindent

Lema\^{\i}tre-Tolman-Bondi (LTB) solutions are used frequently to
describe the collapse or expansion of spherically symmetric
inhomogeneous mass distributions in the Universe. These exact
solutions are obtained in the synchronous gauge where nonlinear
dynamics (with respect to the FLRW background) induce large
deviations from the FLRW metric. In this paper we show explicitly
that this is a gauge artefact (for realistic sub-horizon
inhomogeneities). We write down the nonlinear gauge transformation
from synchronous to Newtonian gauge for a general LTB solution
using the fact that the peculiar velocities are small. In the
latter gauge we recover the solution in the form of a weakly
perturbed FLRW metric that is assumed in standard cosmology.
Furthermore we show how to obtain the LTB solutions directly in
Newtonian gauge and illustrate how the Newtonian approximation
remains valid in the nonlinear regime where cosmological
perturbation theory breaks down. Finally we discuss the
implications of our results for the backreaction scenario.

\end{abstract}

\vskip-19cm  \vskip3in

\end{titlepage}
\setcounter{footnote}{0} \setcounter{page}{1}
\newpage
\baselineskip=20pt





\section{Introduction}
The Friedmann-Lema\^{\i}tre-Roberston-Walker (FLRW) framework is
the cornerstone of modern cosmology. Its key assumption is that
our large scale Universe is isotropic and homogeneous up to small
perturbations. Until not so long ago this was really an
assumption, but fortunately, present observations like the Sloan
Digital Sky Survey show that for distance scales $\gtrsim 100$Mpc
one does indeed seem to find statistical homogeneity
\cite{Hogg:2004vw}\footnote{That is if we don't consider the
possibility of us living at the center of a Hubble-sized void.}.
At the same time however we know that at small scales $\lesssim
10$Mpc our Universe looks like anything but the idealized FLRW
Universe. Nearly all of the matter has clumped into clusters and
galaxies with local density contrasts and dynamics that are
clearly beyond a linear description on a FLRW background. Given
the nonlinear nature of GR, one may wonder then if the FLRW
description of our large scale Universe is really justified. And
if the corresponding scale factor evolves according to the
Friedmann equations, with an effective matter density that is
simply the averaged actual matter density.

This issue was studied in the past but gained much interest over
the last few years with the discovery of the apparent cosmic
acceleration (see \cite{Celerier:2007jc,Rasanen:2006kp} for a
review and an extensive list of references). Within the
conventional FLRW framework this acceleration demands the presence
of a cosmological constant, some other form of dark energy or a
large distance modification of gravity. All cases require the ad
hoc introduction of a new mass scale in the Lagrangian that is
suspiciously of the same magnitude as the present Hubble constant.
In light of this, several papers advocate the so called
backreaction scenario
\cite{Rasanen:2006kp,Rasanen:2008it,Buchert:2007ik,Larena:2008be,Notari:2005xk,Kolb:2005da,Wiltshire:2007jk}.
According to this scenario the conventional FLRW framework is not
justified. The hope is then that a correct averaging of the
inhomogeneities at small distance scales would lead to an
effective energy-momentum tensor to be used in the FLRW
description at large scales, that would obviate the need for dark
energy or modified gravity. A nice feature of this scenario is
that it would naturally solve the coincidence problem (why now?)
since it ties the epoch of cosmic acceleration to the epoch of
nonlinear structure formation.

However, a strong argument against this scenario was raised by
Ishibashi and Wald. In \cite{Ishibashi:2005sj} they simply point
out that, despite the nonlinear dynamics at small scales, the
actual space-time metric of our Universe seems to be very well
approximated by a Newtonianly perturbed FLRW metric at all scales.
By this, they mean a metric of the form \footnote{For simplicity
we are assuming flat FLRW here.} \be ds^2=-\left(1+2\psi \right)
dt^2+a(t)^2\left(1-2\psi\right)(dx^2+dy^2+dz^2) \,,
\label{pertFLRW}\ee with \be |\psi|\ll 1\,\,\,,\,\,\,(\partial_t
\psi)^2\ll
\f{1}{a^2}\partial_i\psi\partial_i\psi\,\,\,,\,\,\,(\partial_i\psi\partial_i\psi)^2\ll
(\partial_i\partial_j\psi)(\partial_i\partial_j\psi)\,\,\,,\ldots\,.\label{condwald}\ee

\textit{If} (\ref{pertFLRW}) is really a good approximation
everywhere to the metric of the Universe, then it is evident that
the backreaction scenario does not work. Indeed, in that case one
can immediately show that the scale factor $a(t)$ in the metric
evolves according to the usual Friedmann equations, up to small
corrections. Of course what we observe are rays of light, not a
metric. But given a metric of the form (\ref{pertFLRW}) one can
resort to a standard lensing analysis to show that the effect of
the small scale inhomogeneities on cosmological observations is
indeed small and can not explain away dark energy. (See
\cite{Holz:2004xx} for instance, for an estimation of the effect
in the context of supernovae observations.)

Now, one thing is the observation that the nonlinear dynamics at
small scales do {\em not preclude} a weakly perturbed FLRW metric.
Another, stronger statement, is that the metric (\ref{pertFLRW})
{\em actually is} a good approximation to the exact metric of the
Universe.  To argue the latter one has to resort to an expansion
in the peculiar velocity $v$ of the full set of Einstein
equations. In this expansion, that lies at the root of the
Newtonian approximation in comoving coordinates, one finds that
(\ref{pertFLRW}) is indeed the leading order approximation to the
metric and that the higher order terms are small. However, this
perturbative argument could in principle still be invalidated by
non-perturbative effects.

In this paper we study this issue in the case of spherically
symmetric space-times, the advantage being that we can use the
exact LTB solutions \cite{LTB} to check the validity of the
Newtonian approximation. Not so surprising, the conclusion will be
that the Newtonian approximation is valid both in the linear and
nonlinear regime. As such this serves as a nice illustration of
the compatibility of a global Newtonianly perturbed FLRW metric
with local nonlinear dynamics. A subtlety will be that the LTB
solutions are formulated in the synchronous gauge. In this gauge,
starting from small initial perturbations, the weak field
description of the metric breaks down at the same time when the
dynamics go nonlinear. In the next section we will show this
explicitly for the LTB solutions, by linearizing the exact
solutions on the FLRW background to recover the standard results
of cosmological perturbation theory. In section 3, we will show
however that this phenomenon is specific to the synchronous gauge.
We will explicitly demonstrate that, for small peculiar
velocities, the corresponding metric in Newtonian gauge remains of
the weakly perturbed FLRW form throughout the entire evolution. In
doing so, we will also verify that the solution that one obtains
in the Newtonian gauge by solving the equations in the Newtonian
approximation, agrees very well with the corresponding exact LTB
solution.

The gauge transformations from synchronous to Newtonian gauge have
been considered before in the context of LTB solutions. A valid
application of the linear transformations was used in
\cite{Biswas:2007gi}, an invalid application can be found in
\cite{Kolb:2008bn}. In \cite{Paranjape:2008ai} the full nonlinear
transformation was performed on a specific LTB solution, obtaining
a different result than ours. We will comment on this further on.

\section{LTB solutions}

The LTB solutions are exact solutions to the Einstein equations
for spherically symmetric distributions of irrotational
pressureless dust, so with energy-momentum tensor
$T_{\mu\nu}(r,t)=\rho u_{\mu}u_{\nu}$. They are obtained in the
synchronous gauge ($g_{00}=-1,g_{0i}=0$) and from now on we will
label the radial and time coordinate in this gauge as
($\t{r},\t{t}$), reserving $(r,t)$ for the Newtonian gauge. Notice
also that for pressureless dust, the synchronous gauge coincides
with the comoving gauge ($u^{\mu}=(1,0,0,0)$). Explicitly the
solutions look like \footnote{We use units such that $c=1$.}:

 \be ds^2=-d\t{t}^2+\f{R'^2(\t{r},\t{t})}{1+2E(\t{r})}d\t{r}^2+R(\t{r},\t{t})^2
d\Omega^2\,,\label{LTBmetric}\ee with\footnote{Some conventions
for the derivatives: the prime stands for a radial and the dot for
a time derivative with respect to the arguments of the function at
hand. Later on we will also use the coordinates $(r,t)$ in
Newtonian gauge, so we will have for instance
$\alpha'=\alpha'(r,t)\equiv \f{\partial \alpha}{\partial r}$ or
$\partial_{r}R=\partial_{r}R(\t{r},\t{t})=\dot{R}\partial_{r}\t{t}+R'\partial_{r}\t{r}\,.$}:
\bea
\f{\dot{R}^2(\t{r},\t{t})}{2}+R(\t{r},\t{t})\ddot{R}(\t{r},\t{t})
=\f{\dot{R}^2(\t{r},\t{t})}{2}-\f{G_NM(\t{r})}{ R(\t{r},\t{t})}&=&E(\t{r})\,,\label{eqY}\\
4\pi\rho(\t{r},\t{t})&=&\f{M'(\t{r})}{R'(\t{r},\t{t})R^2(\t{r},\t{t})}\,\,,\label{rho}\eea
solved by (with $E(\t{r}) \in \mathbb{R}$):
\bea R&=&\f{G_NM}{2E}\left(\cosh \left(u\sqrt{2E}\right) -1\right)\,,\label{R}\\
\t{t}-t_b(\t{r})&=&\f{G_NM}{2E}(\f{\sinh(u\sqrt{2E})}{\sqrt{2E}}-u
)\,.\label{time}\eea

 A particular solution is then specified by some set of
initial conditions \newline
$\{\rho(\t{r},\t{t}_i),R(\t{r},\t{t}_i),\dot{R}(\t{r},\t{t}_i)\}$
at time $\t{t}_i$ that translate to a certain choice for the free
functions $\{E(\t{r}),M(\t{r}),t_b(\t{r})\}$ via eqs.
(\ref{eqY})-(\ref{time}). One of the initial conditions fixes the
residual gauge freedom, we will choose
$R(\t{r},\t{t}_i)=a_i\t{r}$\,\,, where $a_i\equiv a(\t{t}_i)$ is
the scale factor at time $\t{t}_i$ of the FLRW background at
infinity. The two other initial conditions fix the initial density
and {\em peculiar velocity} $v$, where the latter is defined as
($H\equiv \dot{a}/a$): \be v\equiv \dot{R}-HR\,. \label{defv}\ee
Some comments are in order before moving on. For the remainder of
the paper we will only consider solutions that approach the
Einstein- de Sitter solution (flat, matter dominated FLRW, with
$a(t)=\left (t/t_0\right)^{2/3}$\,) \,at infinity, but it's
straightforward to extrapolate our results to situations with
non-flat FLRW backgrounds. One could even add radiation and/or
dark energy for the FLRW background at infinity by taking slightly
modified LTB solutions. Furthermore we will restrict ourselves to
solutions that at some early time were close to the FLRW solution,
in accord with the standard picture of the early Universe. It's
also clear that $v(\t{r},\t{t})$ corresponds to the velocity of
the shell of matter labelled by $\t{r}$ with respect to the FLRW
background. Of course, by definition, the peculiar velocity in
comoving/synchronous gauge is zero. But we will show later on that
$v$ does indeed correspond to the proper peculiar velocity in
Newtonian gauge (at least if $v\ll1$).

Let us now consider some particular solution, describing the
evolution of an initial inhomogeneity, characterized by a velocity
and density profile at some early time $\t{t}_i$: \bea
v_i(\t{r})&\equiv&v(\t{r},\t{t}_i)\,,\\
\rho_i(\t{r})&\equiv&\rho(\t{r},\t{t}_i)\equiv\rho_{
FLRW}(\t{t}_i)(1+\delta(\t{r},\t{t}_i))\equiv
\rho_{FLRW}(\t{t}_i)(1+\delta_i(\t{r})) \,,\eea with  \bea
\delta_i,v_i\ll 1\,\,\,\,\,\,,\,\,\,\lim_{\t{r}\rightarrow 0}
v_i(\t{r})=0\,\,\,\,,\,\,\,\,\lim_{\t{r}\rightarrow \infty}
v_i(\t{r})=0\,\,\,\,,\,\,\,\,
\int_0^\infty\!\!\!d\t{r}\,\t{r}^2\delta_i(\t{r})=0\,.\eea From
eqs. (\ref{eqY})-(\ref{time}) and the first Friedmann equation we
immediately find: \bea
M(\t{r})&=&\f{4\pi\t{r}^3a_i^3}{3}\rho_{FLRW}(\t{t}_i)(1+3I_{\delta_i})\,,\\
E(\t{r})&=&v_iH_ia_i\t{r}-\f{3}{2}(H_ia_i\t{r})^2I_{\delta_i}+\f{v_i^2}{2}\,,\\
t_b(\t{r})&\approx&
\f{1}{H_i}\left(\f{2}{5}\f{E}{(a_iH_i\t{r})^2}+I_{\delta_i}\right)\,,\eea
where we have defined \be I_{\delta_i}(\t{r})\equiv
\f{1}{\t{r}^3}\int_0^{\tilde{r}}\!\!\!dr'\,r'^2\delta_i(r')\,.\ee
It is instructive now to expand the exact solution for
$R(\t{r},\t{t})$ in $\delta_i,v_i$ (treating both quantities as
small parameters of the same order). This is precisely the
expansion that is considered in cosmological perturbation theory
to obtain an approximate solution to the equations of motion for
more generic (non-symmetric) initial conditions. For the case at
hand we find at leading order (keeping the full expression of $ E
\sim v_i \sim \delta_i$): \bea R(\t{r},\t{t})&\approx& \t{r}
a(\t{t})\left(1+\left(\left(\f{\t{t}}{\t{t}_i}\right)^{2/3}\left(\f{2}{5}\f{E}{(a_iH_i\t{r})^2}\right)+
I_{\delta_i}-\f{\t{t}_i}{\t{t}}\left(\f{2}{5}\f{E}{(a_iH_i\t{r})^2}+I_{\delta_i}\right)\right)
+\ldots \right) \,,\nonumber\\ \label{R linear}\,\eea with the
dominant part (for $\t{t}\gg \t{t}_i$) of the higher order terms
going like  \be
\sim\left(\left(\f{\t{t}}{\t{t}_i}\right)^{2/3}\f{E}{(a_iH_i\t{r})^2}\right)^n
\,.\ee

We recognize the first term in the linear term corresponding to
the growing mode, and the last term corresponding to the decaying
mode of cosmological perturbation theory (see for instance
\cite{Ma:1995ey} ). Furthermore, from the behavior of the higher
order terms we see that the expansion breaks down at a time
$t_{nl}$ when \be
\left(\f{t_{nl}}{\t{t}_i}\right)^{2/3}\f{E}{(a_iH_i\t{r})^2}\sim 1
\,.\ee


Now, as we already commented in the introduction, one thing is the
breakdown of the expansion in $\delta_i$, another thing is the
breakdown of the metric description in terms of a weakly perturbed
FLRW metric. But as one can see from (\ref{R linear}), the
breakdown of the former automatically implies the breakdown of the
latter for the synchronous gauge solution that we are considering.
Indeed we find \be
\f{|R(\t{r},\t{t})-a(\t{t})\t{r}|}{a(\t{t})\t{r}}\sim 1\,,\ee at
$\t{t}\sim t_{nl}$. In addition one can see from the exact
expression for the matter density (\ref{rho}), that the density
contrast also becomes large at the same time:
$\delta(\t{r},t_{nl})\sim 1$.

Notice that the converse is not necessarily true, one can have
solutions with shell-crossing singularities
($g_{\t{r}\t{r}}=R'=0$), for which the description of the metric
in terms of a weakly perturbed FLRW clearly breaks down, that are
perfectly well described by the perturbative expansion in
$\delta_i$ for times arbitrary close to the time of
shell-crossing. The "small u expansion" of \cite{Biswas:2006ub}
uses precisely this feature in combination with the exact
expression (\ref{rho}) to describe the density.

\section{From synchronous to Newtonian gauge (and back)}
\subsection{The linear case}
We saw in the previous section how the description of the LTB
metric as a weakly perturbed FLRW metric in synchronous gauge,
breaks down at the same time when the dynamics on the FLRW
background become nonlinear. Let us now see what happens in
Newtonian (or Poisson) gauge. In spherically symmetric situations,
given the standard angular coordinates, this gauge is specified by
the conditions $g_{rt}=0$ and $g_{rr}r^2=g_{\theta\theta}$. So we
want to find the coordinate transformation
$(\t{r},\t{t})\rightarrow(r,t) $ that changes the metric to: \be
ds^2= -dt^2\left(1+2\phi(r,t)\right)+ a^2(
t)\left(dr^2+r^2d\Omega^2\right)\left(1-2\psi(r,t)\right)\,.\label{newmetric}\ee

At times $t\ll t_{nl}$, when the dynamics are still linear, the
coordinate transformation will also be linear\footnote{By a linear
coordinate transformation we mean that $\t{t}=t+\alpha$ and $
\t{r}=r+\beta\,$ with $\alpha',\dot{\alpha},\beta',\dot{\beta}\ll
1$.} and one can resort directly to the machinery developed in the
context of cosmological perturbation theory to obtain the
potentials \cite{Ma:1995ey}. This was done for instance in
\cite{Biswas:2007gi} and for our conventions we find in a
completely similar fashion that the resulting potentials read: \be
\phi\approx \psi +\mathcal{O}(\psi)^2\approx
-\f{3}{5}\int\!\!\!dr\f{E(r)}{r}\,\,+\,\,\f{3}{2}\left(\f{t_i}{t}\right)^{5/3}\int\!\!\!dr
\left( \f{2}{5}\f{E(r)}{r}+r(a_i
H_i)^2I_{\delta_i}(r)\right)\,.\ee One can again recognize the
growing mode and the decaying mode. As we should, we find the
former to result in a constant term for the potential in Newtonian
gauge. This, in contrast with the situation for synchronous gauge,
implies that the metric keeps its form of a weakly perturbed FLRW
($\psi\approx\phi\ll 1$) up to the time $t_{nl}$, when the
dynamics become nonlinear. Of course one can not resort to
cosmological perturbation theory anymore to argue for the
smallness of the fields at later times, when the fields and the
actual density contrast $\delta$, become nonlinear functions of
$\delta_i$. Still, from the argument of Ishibashi and Wald, we
might expect the potential ($\psi\approx\phi$) to remain small for
sub-horizon inhomogeneities. As we will show explicitly in the
next subsection, this is indeed the case for the LTB solutions in
Newtonian gauge, provided that the peculiar velocities remain
small. For the LTB solutions this will hold as long as we stop the
evolution well before a central singularity develops. In reality,
when we consider the collapse of a cluster for instance, the
(effective) pressure that arises during the virialization, will
halt the collapse.

\subsection{The nonlinear case}

We could proceed now in the same way as we did for the linear
case. So, starting from the LTB solution we could look for the
coordinate transformation $(\t{r},\t{t})\rightarrow (r,t)$, which
is now nonlinear, to obtain the metric in the Newtonian gauge.
This strategy was used in \cite{Paranjape:2008ai}. But we find it
more instructive go the other way. We will start from the
equations in Newtonian gauge for spherically symmetric situations,
with metric (\ref{newmetric}), showing explicitly how the
expansion in the peculiar velocity $v$ gives rise to the familiar
equations of the Newtonian approximation to GR. Then we will
perform the coordinate transformation to synchronous gauge,
demonstrating explicitly that the solutions obtained in the
Newtonian approximation are indeed good approximations to the
exact LTB solutions.

First we should explain at last what we mean exactly by the
expansion in $v$. From Newtonian physics one gets the following
order of magnitude estimates in the case of a non-virialized
system with density contrast $\delta$ and distance scale
$L$\cite{peebles}: \be \psi\sim\phi\sim \delta
(HL)^2\,\,\,,\,\,\,\psi'\sim\delta H^2L\,\,\,,\,\,\,v\sim \delta
HL\,\,\,,\,\,\, \dot{\psi}\sim \psi
\f{v}{L}\,\,\,,\,\,\,\dot{v}\sim \f{v^2}{L}\,\,\,,\ldots
\,.\label{newtoncond}\ee When assigning a power of $v$ to a
particular term in the expansion, we will use the order of
magnitude estimates above, with $\delta\sim 1$. This ensures that
the expansion remains valid in the nonlinear regime $\delta\gtrsim
1$.

Let us now apply this expansion to the full set of Einstein
equations. In the spherically symmetric case that we are
considering, there are four independent equations: \be
G_{tt}\approx 8 \pi G_N \rho\,\,,\,\,G_{tr}\approx -8 \pi G_N \rho
v a\,\,,\,\,G_{rr}\approx 8 \pi G_N\rho v^2
a^2\,\,,\,\,G_{\varphi\varphi}=\sin^2\!\theta \,G_{\theta
\theta}=0\,, \ee for the perfect dust energy-momentum tensor
$T_{\mu\nu}=\rho u_\mu u_\nu$, with $u^{\mu}\approx (1,v/a,0,0)$.
We easily get the fields $\phi$ and $\psi$ from the
($\theta,\theta$) and ($t,t$) equations. The former

\be
G_{\theta\theta}=r^2(\phi''-\psi''+\f{1}{r}\phi'-\f{1}{r}\psi')\,\,+
\mathcal{O}(v^4)=0\,, \ee immediately tells us that
$\psi=\phi+\mathcal{O}(v^4)$.\footnote{We are imposing the
boundary conditions $\phi,\psi,\phi',\psi'\rightarrow 0$ for
$r\rightarrow \infty$, to match the FLRW background at infinity.}
The latter then reduces to the Poisson equation (using the FLRW
equation for the background), with the potential sourced by the
density contrast: \be \nabla^2\phi=4\pi G_N a^2\rho_{FLRW} \delta
\,\,+H^2\mathcal{O}(v^2)\,.\label{poisson} \ee The metric
(\ref{newmetric}) will then solve the other two Einstein equations
if the energy-momentum tensor is conserved. That is if $\rho$ and
$v$ obey the familiar ideal fluid equations in comoving
coordinates (for zero pressure) \cite{peebles}: \bea
\nabla_{\mu}T^{\mu t}= \left(\dot{\rho}\,\,+3H\rho+\f{1}{a}(\rho
v)'+\f{2}{ra}\rho v\right) +
\f{H^2}{G_NL}\mathcal{O}(v^3)&=&0\,,\label{cont}\\
\f{a}{\rho}\left(\nabla_{\mu}T^{\mu r} - \f{v}{a}
\nabla_{\mu}T^{\mu t}\right)=\left(\dot{v}+\f{v v'}{a}+ H
v+\f{\phi'}{a}\right)+\f{1}{L}\mathcal{O}(v^4)&=&0\,.\label{euler}\eea

The solution in the Newtonian approximation can now be obtained by
solving the eqs. (\ref{poisson})-(\ref{euler}) at leading order,
for a particular initial velocity and density profile
$v_i(r),\delta_i(r)\ll 1$ at some early time $t_i$. One can verify
that for these solutions, the order of magnitude estimates
(\ref{newtoncond}) are correct, so the higher order terms in the
expansion will indeed be suppressed, both in the linear
($\delta\ll 1$) and nonlinear ($\delta\gtrsim 1$) regime as long
as $v$ remains small. This type of argument on the validity of the
Newtonian approximation could be criticized for being circular. In
a sense we are using the Newtonian approximation to justify
itself. But as we will now show by going to the synchronous gauge,
the approximate solution obtained in the Newtonian gauge, is
indeed a very good approximation to the exact LTB solution.

We define the coordinate transformation by: \bea
R(\t{r},\t{t})^2&=&a(t)^2r^2(1-2\psi(r,t))\,,\label{Y}\\
\t{t}&=&t+\alpha(r,t) \,,\label{t}\eea for some functions
$R(\t{r},\t{t})$ and $\alpha(r,t)$ that will be determined from
the synchronous gauge conditions $g_{\t{t}\t{r}}=0$ and
$g_{\t{t}\t{t}}=-1$. We will do this in the same expansion in $v$
that we used to obtain the Newtonian approximation, retaining only
the terms up to $\mathcal{O}(v^2)$. Immediately we can anticipate
that \be \alpha'\approx-a v\,, \label{defalpha}\ee from the
condition that $u^{\t{r}}$ should be zero in the
synchronous/comoving gauge. Applying the coordinate transformation
on the metric, with the use of \bea
d\t{t}&=&dt\left(1+\dot{\alpha}\right)+dr\alpha'\,,\label{dt}\\
R'd\t{r}&\approx&dt\left(\dot{a}r-\dot{R}\right)+dr\left(a(1-\psi)-ar\psi'-\dot{R}\alpha'\right)\,,\label{dY}\eea
we then find \be g_{\t{t}\t{r}}\approx
\f{\alpha'R'}{a}+R'(\dot{R}-\dot{a}r)=0\,,\ee for \be
\dot{R}-\dot{a}r \approx\dot{R}-HR\approx -\f{\alpha'}{a}
\,\left(\,\approx v\right)\,.\label{vapprox}\ee Employing this
expression it is then straightforward to show that $g_{\t{t}\t{t}}
\approx -1$,  if $v$ obeys the Euler equation (\ref{euler}). So
the transformation (\ref{Y})-(\ref{t}) with $\alpha$ obeying
(\ref{defalpha}) and $R$ obeying (\ref{vapprox}) indeed takes us
to the synchronous gauge.

We will now show that $R$ approximately solves the LTB equation if
$v,\rho$ and $\phi$ solve the eqs. (\ref{poisson})-(\ref{euler}),
that we obtained in the Newtonian approximation. Let us first look
at the continuity equation (\ref{cont}) and show that it is
equivalent to the LTB expression (\ref{rho}) for the matter
density. Expressing the $(r,t)$ derivatives in terms of
$(\t{r},\t{t})$ derivatives through eqs. (\ref{dt}) and
(\ref{dY}), this equation becomes (again keeping the appropriate
powers of $v$): \be
\partial_{\tilde{t}}\rho+ 3H \rho
+\rho\f{\partial_{\t{r}}v}{R'}+\f{2}{ra}\rho v\approx 0 \,.\ee
Using the expressions (\ref{Y}) and (\ref{vapprox}) for $R$ and
$v$, this reduces to \be \partial_{\tilde{t}}\rho+
\rho\left(\f{\dot R'}{R'}+ 2 \f{\dot{R}}{R}\right)\approx 0 \,,\ee
solved by the LTB expression (\ref{rho}) for $\rho$.

Employing this LTB expression for $\rho$ in the Poisson equation
(\ref{poisson}) for $\phi$, we find in a similar way that: \bea
\phi'&\approx& \f{a^2}{r^2}\left(\int\!
\!dr\,r^2\left(\f{G_NM'(\t{r})}{R'R^2}-\f{3}{2}H^2\right)\right)
\nonumber\\&\approx&\f{a}{
R^2}\left(\int\!\!d\t{r}\,G_NM'(\t{r})\right)-\f{a H^2
R}{2} \,\nonumber\\
&\approx&a\left(\f{G_N M}{R^2}-\f{H^2 R}{2}\right)\,.\eea If we
now use this expression for $\phi'$ in combination with
$a(t)=(t/t_0)^{2/3}$, we finally find that the Euler equation
(\ref{euler}) reduces to:

\be \ddot{R}+\f{G_N
M(\t{r})}{R^2}=\f{1}{\dot{R}}\f{\partial}{\partial
\t{t}}\left(\f{\dot{R}^2}{2}-\f{G_N M(\t{r})}{R} \right)\approx
0,\ee which is now indeed solved by the LTB equation (\ref{eqY}).
So we have demonstrated that for a solution obtained in the
Newtonian approximation one finds
$g_{\theta\theta}=R(\t{r},\t{t})^2$ in the synchronous gauge, with
$R$ an approximate solution of the exact LTB equations.  Keeping
track of the omitted terms in our expansion one can show that the
approximation holds up to terms $\sim v^4$. As for the initial
conditions, it is easy to see that the initial density and
velocity profile at time $t_i$ in the Newtonian gauge, translate
to (approximately) the same initial conditions at time $\t{t}_i$
for the LTB solutions: $\delta_i(r),v_i(r)\approx
\delta_i(\t{r}),v_i(\t{r})$. That is if $r\approx \t{r}$ around
the time $t_i$, which is true if we fix the residual gauge degree
of freedom by the condition $R(\t{r},\t{t_i})=a_i \t{r}$, as we
did in the previous section.

A nice cross check of our derivation is provided by the
calculation of $g_{\t{r}\t{r}}$. From the exact LTB solution we
know that we should find \be g_{\t{r}\t{r}}=\f{{R'}^2}{1+2E}
\,,\ee whereas from the coordinate transformation
(\ref{dt})-(\ref{dY}) we find, \be g_{\t{r}\t{r}}\approx
R'^2\left(1+2\left(\f{\dot{a}}{a}\alpha'r+\psi'
r\right)-\f{\alpha'^2}{a^2}\right)\,.\ee Using the expressions for
$R$ (\ref{Y}) and $\alpha$ (\ref{vapprox}) this indeed reduces to
\bea
g_{\t{r}\t{r}}&\approx&R'^2\left(1-\dot{R}^2+\f{2G_NM}{R}\right)\nonumber\\
&\approx&\f{R'^2}{1+2E}\,, \eea where on the last line we have
used the LTB equation (\ref{eqY}) and $E\ll 1$.

To recapitulate, we have demonstrated that the coordinate
transformation, implicitly defined by:

\bea
R(\t{r},\t{t})^2&=&a(t)^2r^2(1-2\psi(r,t))\,,\label{trans1}\\
\t{t}&=&t+a(t)\int^{\infty}_{r}\!\!dr' v(r',t)\label{trans2}
\,,\eea with \be \dot{R}(\t{r},\t{t})-H(\t{t})R(\t{r},\t{t})=
v(r,t)\,,\ee takes the metric (\ref{newmetric}) in Newtonian gauge
to synchronous gauge, provided $v$ is small.  In the synchronous
gauge we recover the metric of the LTB form (\ref{LTBmetric}),
with $R$ and $\rho$ approximately obeying the LTB equations
(\ref{eqY}) and (\ref{rho}) in synchronous coordinates, if
$\rho,v,\phi\approx \psi$ are solutions of the Poisson, continuity
and Euler equations in the Newtonian coordinates.

It's clear that we can use our results also in the other
direction. Namely if we start from the exact LTB solution in
synchronous gauge, the transformation
(\ref{trans1})-(\ref{trans2}) will take the metric to a
Newtonianly perturbed FLRW metric with $\rho,\psi,v$ solving the
Poisson, continuity and Euler equation in Newtonian coordinates at
leading order in $v$.  We should stress (at the risk of being
repetitive) that our analysis is valid both in the linear and
nonlinear regime, as long as the peculiar velocities remain small.
Notice also that the conditions on $\psi$ (\ref{condwald}) imposed
by Ishibashi and Wald are automatically satisfied, again if $v$ is
small.

As we mentioned, in \cite{Paranjape:2008ai} a similar analysis was
performed for the specific case of zero initial velocity and an
initial density profile describing a constant over-density in the
core surrounded by a finite region with a constant under-density.
Surprisingly, the authors found a considerable difference for the
fields $\psi$ and $\phi$ when the dynamics go nonlinear, seemingly
in conflict with the Newtonian approximation and with our results.
However, it turns out that the reason for the discrepancy lies in
a calculational error \cite{erroraseem}.

\section{Conclusions}
In this paper we have demonstrated explicitly how one can recover
the exact LTB solutions in synchronous gauge, from the
corresponding solutions obtained in the Newtonian approximation.
This was done by applying the full nonlinear coordinate
transformation from Newtonian to synchronous gauge, using the fact
that the peculiar velocities remain small, which is the case for
realistic sub-horizon inhomogeneities.

As we have illustrated, in their original form the LTB solutions
display a breakdown of the weak field description of the metric,
when the dynamics go nonlinear with respect to the FLRW
background. This has been used to argue for the backreaction
scenario by Rasanen \cite{Rasanen:2006kp,Rasanen:2008it} for
instance. Indeed, once the weak field description breaks down, one
would not a expect a priori that a Universe full of collapsing
structures and expanding voids would still give a FLRW metric {\em
on the average} with a normal evolution of the scale factor.
However, as we have shown, the breakdown of the weak field
description of the metric is in fact specific to the synchronous
gauge. In Newtonian gauge, the LTB solutions are perfectly well
described by a Newtonianly perturbed FLRW metric and it becomes
straightforward to show that both the metric and the dynamics will
on the average behave as a conventional FLRW Universe (up to small
corrections).

LTB solutions have been used also in the literature to model the
effect of inhomogeneities on light propagation in the so called
Swiss cheese models \cite{Brouzakis:2007zi}. Again, given the
breakdown of the weak field description, one would not expect a
priori to approximately recover the standard FLRW
luminosity-distance redshift relation, when averaging over all
directions. Yet this is precisely what is found and is again easy
to understand from the weak field expansion in Newtonian gauge, as
was demonstrated for LTB solutions in the linear regime in
\cite{Biswas:2007gi}. In a forthcoming publication we will use the
full transformation (\ref{trans1})-(\ref{trans2}) to analyze the
Swiss cheese model of \cite{Marra:2007pm} in the nonlinear regime.
We will find that the metric indeed reduces to a weakly perturbed
FLRW metric in the Newtonian gauge, which was used implicitly in
the recent paper \cite{Vanderveld:2008vi}.

All this illustrates the strength of the Newtonian approximation
{\em in comoving coordinates}\footnote{As opposed to the Newtonian
approximation in physical coordinates, which breaks down at
cosmological distance scales.} in justifying the conventional FLRW
framework. And it is clear that any serious backreaction scenario
should explain where and how the Newtonian approximation breaks
down. Actually we know that the Newtonian approximation breaks
down in the vicinity of black holes, and to our knowledge it has
not been fully demonstrated yet that the corresponding
backreaction is negligible. Notice also that in this paper we have
only strictly proven the validity of the Newtonian approximation
for spherically symmetric space-times, that approach the FLRW
solution at infinity. But as we commented in the introduction, for
more general space-times one can show that the Newtonian
approximation is at least self consistent.

Finally we should comment on the local void scenario, which is
another approach that involves the use of inhomogeneity in trying
to dispose of dark energy. In this scenario one puts us near the
center of a large void, typically described by an LTB solution.
The mismatch between the local and global expansion can then
explain the supernovae data
\cite{Tomita:2001gh,Alnes:2005rw,Vanderveld:2006rb}. However, to
explain other data sets, like those on the CMB and the large scale
structure, one has to introduce additional features in the
primordial power spectrum and the matter composition of the
Universe \cite{Sarkar:2007cx,Alexander:2007xx}. Another contrast
with the backreaction scenario is that this scenario can be
perfectly well described in a weak field description, with the
local Hubble flow encoded in the Doppler term (see for instance
\cite{Biswas:2007gi}). Indeed, a good fit to the supernovae data
requires a shift in the local Hubble parameter of order 10\% with
respect to the global value that is recovered at a distance
$r\sim$ 1Gpc. This translates to a maximal peculiar velocity
$v\sim \Delta H \,r \lesssim 0.05$ which is still rather small and
therefore validates the Newtonian approximation. This issue was
studied by the authors of \cite{Kolb:2008bn}, for the type of void
proposed in \cite{Alnes:2005rw}. For this void, the present
peculiar velocity is indeed small, and since the solution is in
the nonlinear regime, one needs the full nonlinear transformation
(\ref{trans1})-(\ref{trans2}) to recover the metric as a Newtonian
perturbation of the FLRW metric. Also, as was noticed in
\cite{Kolb:2008bn}, a specific feature of this void solution is
that the peculiar velocities become large ($v\gtrsim 1$) in the
past, at redshifts $z\gtrsim 500$.  At that time the void
represents a nonlinear super-horizon density fluctuation and the
weak field description in Newtonian gauge will of course break
down.

\section*{Acknowledgments}
It's a pleasure to thank Valerio Marra, Aseem Paranjape, T.P.
Singh, Syksy Rasanen and Thomas Buchert for stimulating
discussions. I am supported by a postdoctoral grant of the Fund
for Scientific Research-Flanders (Belgium).


\begin{thebibliography}{1}
\bibitem{Hogg:2004vw}
  D.~W.~Hogg, D.~J.~Eisenstein, M.~R.~Blanton, N.~A.~Bahcall, J.~Brinkmann, J.~E.~Gunn and D.~P.~Schneider,
  Astrophys.\ J.\  {\bf 624} (2005) 54
  [arXiv:astro-ph/0411197].

  \bibitem{Celerier:2007jc}
  M.~N.~Celerier,
  arXiv:astro-ph/0702416.


\bibitem{Rasanen:2006kp}
  S.~Rasanen,
  JCAP {\bf 0611} (2006) 003
  [arXiv:astro-ph/0607626].


\bibitem{Rasanen:2008it}
  S.~Rasanen,
  JCAP {\bf 0804} (2008) 026
  [arXiv:0801.2692 [astro-ph]].



\bibitem{Buchert:2007ik}
  T.~Buchert,
  Gen.\ Rel.\ Grav.\  {\bf 40} (2008) 467
  [arXiv:0707.2153 [gr-qc]].


\bibitem{Larena:2008be}
  J.~Larena, J.~M.~Alimi, T.~Buchert, M.~Kunz and P.~S.~Corasaniti,
  arXiv:0808.1161 [astro-ph];

\bibitem{Notari:2005xk}
  A.~Notari,
  Mod.\ Phys.\ Lett.\  A {\bf 21} (2006) 2997
  [arXiv:astro-ph/0503715];
\bibitem{Kolb:2005da}
  E.~W.~Kolb, S.~Matarrese and A.~Riotto,
  New J.\ Phys.\  {\bf 8} (2006) 322
  [arXiv:astro-ph/0506534];
\bibitem{Wiltshire:2007jk}
  D.~L.~Wiltshire,
  New J.\ Phys.\  {\bf 9} (2007) 377
  [arXiv:gr-qc/0702082].


\bibitem{Ishibashi:2005sj}
  A.~Ishibashi and R.~M.~Wald,
  Class.\ Quant.\ Grav.\  {\bf 23} (2006) 235
  [arXiv:gr-qc/0509108].

\bibitem{Holz:2004xx}
  D.~E.~Holz and E.~V.~Linder,
  Astrophys.\ J.\  {\bf 631} (2005) 678
  [arXiv:astro-ph/0412173].


\bibitem{LTB} G. Lema\^{\i}tre, Ann. soc. Sci. Bruxelles Ser.1, A53, 51, 1933;
              R. C. Tolman,Proc. Nat1. Acad. Sci. U.S.A. 20,410, 1934;
              H. Bondi, Mon. Not. R. Astron. Soc., 107, 343, 1947).

\bibitem{Biswas:2007gi}
  T.~Biswas and A.~Notari,
  JCAP {\bf 0806} (2008) 021
  [arXiv:astro-ph/0702555].

\bibitem{Kolb:2008bn}
  E.~W.~Kolb, V.~Marra and S.~Matarrese,
  arXiv:0807.0401 [astro-ph].

\bibitem{Paranjape:2008ai}
  A.~Paranjape and T.~P.~Singh,
  JCAP {\bf 0803} (2008) 023
  [arXiv:0801.1546 [astro-ph]].


\bibitem{Ma:1995ey}
  C.~P.~Ma and E.~Bertschinger,
  Astrophys.\ J. {\bf 4 55}  (1995) 7
  [arXiv:astro-ph/9506072].

\bibitem{Biswas:2006ub}
  T.~Biswas, R.~Mansouri and A.~Notari,
  JCAP {\bf 0712} (2007) 017
  [arXiv:astro-ph/0606703].

  \bibitem{peebles}
P.~J.~E. Peebles, ``The Large Scale structure of the Universe,''
Princeton University Press (1980).



\bibitem{erroraseem} In  \cite{Paranjape:2008ai} the term
$\f{{{\xi^{0}}'}^2}{2R'R}$ is missing on the right-hand side of
Eq. (35). Taking this term into account one finds an additional
term $-\f{1}{4}(a\t{v})^2$ in the field $\tilde{\psi}$, making
$A+\t{\psi}=0$, in agreement with our results.

\bibitem{Brouzakis:2007zi}
  N.~Brouzakis, N.~Tetradis and E.~Tzavara,
  JCAP {\bf 0804} (2008) 008
  [arXiv:astro-ph/0703586];
  N.~Brouzakis and N.~Tetradis,
  Phys.\ Lett.\  B {\bf 665} (2008) 344
  [arXiv:0802.0859 [astro-ph]];
  N.~Brouzakis, N.~Tetradis and E.~Tzavara,
  JCAP {\bf 0702} (2007) 013
  [arXiv:astro-ph/0612179].


\bibitem{Marra:2007pm}
  V.~Marra, E.~W.~Kolb, S.~Matarrese and A.~Riotto,
  Phys.\ Rev.\  D {\bf 76} (2007) 123004
  [arXiv:0708.3622 [astro-ph]].

\bibitem{Vanderveld:2008vi}
  R.~A.~Vanderveld, E.~E.~Flanagan and I.~Wasserman,
  arXiv:0808.1080 [astro-ph].










\bibitem{Tomita:2001gh}
  K.~Tomita,
  Prog.\ Theor.\ Phys.\  {\bf 106} (2001) 929
  [arXiv:astro-ph/0104141].

\bibitem{Alnes:2005rw}
  H.~Alnes, M.~Amarzguioui and O.~Gron,
  Phys.\ Rev.\  D {\bf 73} (2006) 083519
  [arXiv:astro-ph/0512006].

\bibitem{Vanderveld:2006rb}
  R.~A.~Vanderveld, E.~E.~Flanagan and I.~Wasserman,
  Phys.\ Rev.\  D {\bf 74} (2006) 023506
  [arXiv:astro-ph/0602476].

\bibitem{Sarkar:2007cx}
  S.~Sarkar,
  Gen.\ Rel.\ Grav.\  {\bf 40} (2008) 269
  [arXiv:0710.5307 [astro-ph]].

\bibitem{Alexander:2007xx}
  S.~Alexander, T.~Biswas, A.~Notari and D.~Vaid,
  arXiv:0712.0370 [astro-ph].


























\end{thebibliography}
\end{document}